\documentclass[aps,reprint,prl,superscriptaddress,showpacs,floats,floatfix]{revtex4-1}
\usepackage{graphicx}
\usepackage{amsmath,graphicx,dcolumn}
\usepackage{hyperref}
\usepackage[usenames]{color}
\usepackage{datetime}
\usepackage{color}
\usepackage{textcomp}
\usepackage{ulem}

\newcommand*{\CPI}{CePt$_2$In$_7$}

\usepackage{graphicx}

\begin{document}

\title{Crystal electric field splitting and $f$-electron hybridization in heavy fermion {\CPI}}

\author{Yu-Xia Duan}
\affiliation{School of Physics and Electronics, Central South University, Changsha 410083, Hunan, Peoples Republic of China}

\author{Cheng Zhang}
\affiliation{School of Physics and Electronics, Central South University, Changsha 410083, Hunan, Peoples Republic of China}

\author{J\'{a}n Rusz}
\affiliation{Department of Physics and Astronomy, Uppsala University, Box 516, S-75120 Uppsala, Sweden}

\author{Peter M. Oppeneer}
\affiliation{Department of Physics and Astronomy, Uppsala University, Box 516, S-75120 Uppsala, Sweden}

\author{Tomasz Durakiewicz}
\affiliation{Institute of Physics, Maria Curie Sklodowska University, 20-031 Lublin, Poland}

\author{Yasmine Sassa}
\affiliation{Department of Physics and Astronomy, Uppsala University, Box 516, S-75120 Uppsala, Sweden}
\affiliation{Department of Physics, Chalmers University of Technology, 41296 G\"{o}teborg, Sweden}

\author{Oscar Tjernberg}
\affiliation{Department of Applied Physics, KTH Royal Institute of Technology, Electrum 229, SE-16440, Stockholm, Kista, Sweden}

\author{Martin M{\aa}nsson}
\affiliation{Department of Applied Physics, KTH Royal Institute of Technology, Electrum 229, SE-16440, Stockholm, Kista, Sweden}

\author{Magnus H. Berntsen}
\affiliation{Department of Applied Physics, KTH Royal Institute of Technology, Electrum 229, SE-16440, Stockholm, Kista, Sweden}

\author{Fan-Ying Wu}
\affiliation{School of Physics and Electronics, Central South University, Changsha 410083, Hunan, Peoples Republic of China}

\author{Yin-Zou Zhao}
\affiliation{School of Physics and Electronics, Central South University, Changsha 410083, Hunan, Peoples Republic of China}

\author{Jiao-Jiao Song}
\affiliation{School of Physics and Electronics, Central South University, Changsha 410083, Hunan, Peoples Republic of China}

\author{Qi-Yi Wu}
\affiliation{School of Physics and Electronics, Central South University, Changsha 410083, Hunan, Peoples Republic of China}

\author{Yang Luo}
\affiliation{School of Physics and Electronics, Central South University, Changsha 410083, Hunan, Peoples Republic of China}

\author{Eric D. Bauer}
\affiliation{Condensed Matter and Magnet Science Group, Los Alamos National Laboratory, Los Alamos, New Mexico 87545, USA}

\author{Joe D. Thompson}
\affiliation{Condensed Matter and Magnet Science Group, Los Alamos National Laboratory, Los Alamos, New Mexico 87545, USA}

\author{Jian-Qiao Meng}
\email{Corresponding author: jqmeng@csu.edu.cn}\affiliation{School of Physics and Electronics, Central South University, Changsha 410083, Hunan, Peoples Republic of China}
\affiliation{Synergetic Innovation Center for Quantum Effects and Applications (SICQEA), Hunan Normal University, Changsha 410081, Peoples Republic of China}

\date{\today}

\begin{abstract}
We use high-resolution angle-resolved photoemission spectroscopy to investigate the electronic structure of the antiferromagnetic heavy fermion compound {\CPI}, which is a member of the CeIn$_3$-derived heavy fermion material family. Weak hybridization among 4$f$ electron states and conduction bands was identified in {\CPI} at low temperature much weaker than that in the other heavy fermion compounds like CeIrIn$_5$ and CeRhIn$_5$. The Ce 4$f$ spectrum shows fine structures near the Fermi energy, reflecting the crystal electric field splitting of the $4f_{5/2}^1$ and $4f_{7/2}^1$ states. Also, we find that the Fermi surface has a strongly three-dimensional topology, in agreement with density-functional theory calculations.
\end{abstract}

\maketitle

The physics underlying the formation of superconductivity from a coherent heavy fermion (HF) state has persisted as a central mystery despite more than four decades of intensive experimental and theoretical study \cite{Pfleiderer2009, Taillefer2013}. HF superconductivity, similar to traditional Bardeen-Cooper-Schrieffer (BCS) superconductivity, is more likely to occur in compounds with particular crystal structures \cite{Heffner1996, Petrovic2001}, and for magnetically mediated superconductivity, quasi-two-dimensional (2D) compounds are likely to have a higher transition temperature than a three-dimensional counterpart \cite{Monthoux2001}. This expectation appears to be borne out in a family of tetragonal HF compounds Ce$_m$$M$$_n$In$_{3m+2n}$ ($M$ = Co, Rh, and Ir) in which the quasi-2D members with $m$=1, $n$=1 have $T$$_c$'s \cite{JDThompson2001A} up to an order of magnitude higher than the maximum pressure-induced $T$$_c$ of 0.25 K  found in cubic building block CeIn$_3$ \cite{IRWalker1997}. Recent polarized soft x-ray absorption and nonresonant inelastic x-ray scattering  experiments find that, in addition to the crystal structure, details of anisotropic hybridization of $f$ and itinerant electrons play a nontrivial role in determining the ground state of the $m$=1, $n$=1 family members \cite{TWillers2015, MSundermann2019}. Understanding how superconductivity originates from a strongly correlated ground state which must be associated with structure, dimensionality, and orbital anisotropy requires detailed high-resolution techniques that allow studying quantitatively a system's electronic structure.

Low-dimensional Ce-based systems hold a promise for the successful angle-resolved photoemission spectroscopy (ARPES) exploration of the HF electronic structure with high momentum and energy resolution. Most prior ARPES investigations of Ce-based HFs have been done on 3D systems but significant $k_z$ broadening \cite{JDDenlinge2001, RJiang2015, Fujimori2003, Fujimori2006, QYChen2018A} limits the intrinsic accuracy of 3D band measurement. Low photon energies produce high-energy and in-plane momentum resolutions; however, the 4$f$ signal is quite low and sits atop a relatively large background. The part of the reciprocal-space spectrum where the nearly flat 4$f$ bands, split by spin-orbit interaction, intersect with conduction bands is the site of interest for detailed inquiry. A Ce-based HF system, as close to 2D as possible, is needed to extract this information and that would provide a level of detail comparable to what is provided by ARPES measurements on the cuprate high-temperature superconductors.

The HF compound, {\CPI}, is an antiferromagnetic superconductor identified in 2008 \cite{Kurenbaeva2008}. It belongs to the Ce$_m$$M$$_n$In$_{3m+2n}$($m$=1, $n$=2) family. {\CPI} bulk superconductivity can be induced by applying pressure, reaching a maximum of $T_c$ = 2.1 K near 3.5 GPa \cite{EDBauer2010A}. Compared to other family members, such as CeIn$_3$ and Ce$M$In$_5$, the distance between adjacent CeIn$_3$ blocks in {\CPI} increases greatly and the $f$/$spd$ hybridization decreases \cite{EDBauer2010A, VASidorov2013}, making it toward the 2D limit in CeIn$_3$-derived HF material family \cite{EDBauer2010B, MMAltarawneh2011}. The crystal structure of {\CPI}, however, is body-centered tetragonal ($I$4/$mmm$) in contrast to the primitive tetragonal structure ($P$4/$mmm$) adopted by the $m$=1, $n$=1 family members \cite{Klimczuk2014}. Though the separation between CeIn$_3$ units is greater in {\CPI}, promoting a more 2D-like band character, the insertion of a PtIn$_2$ layer between CeIn$_3$ units modifies the $f$/$d$ hybridization pathway and produces a notable 3D character that is absent in the simpler $m$=1,$n$=1 members \cite{Klimczuk2014, MMAltarawneh2011, BShen2017}.

\begin{figure}[t!]
\vspace*{-0.2cm}
\begin{center}
\includegraphics[width=0.95\columnwidth,angle=0]{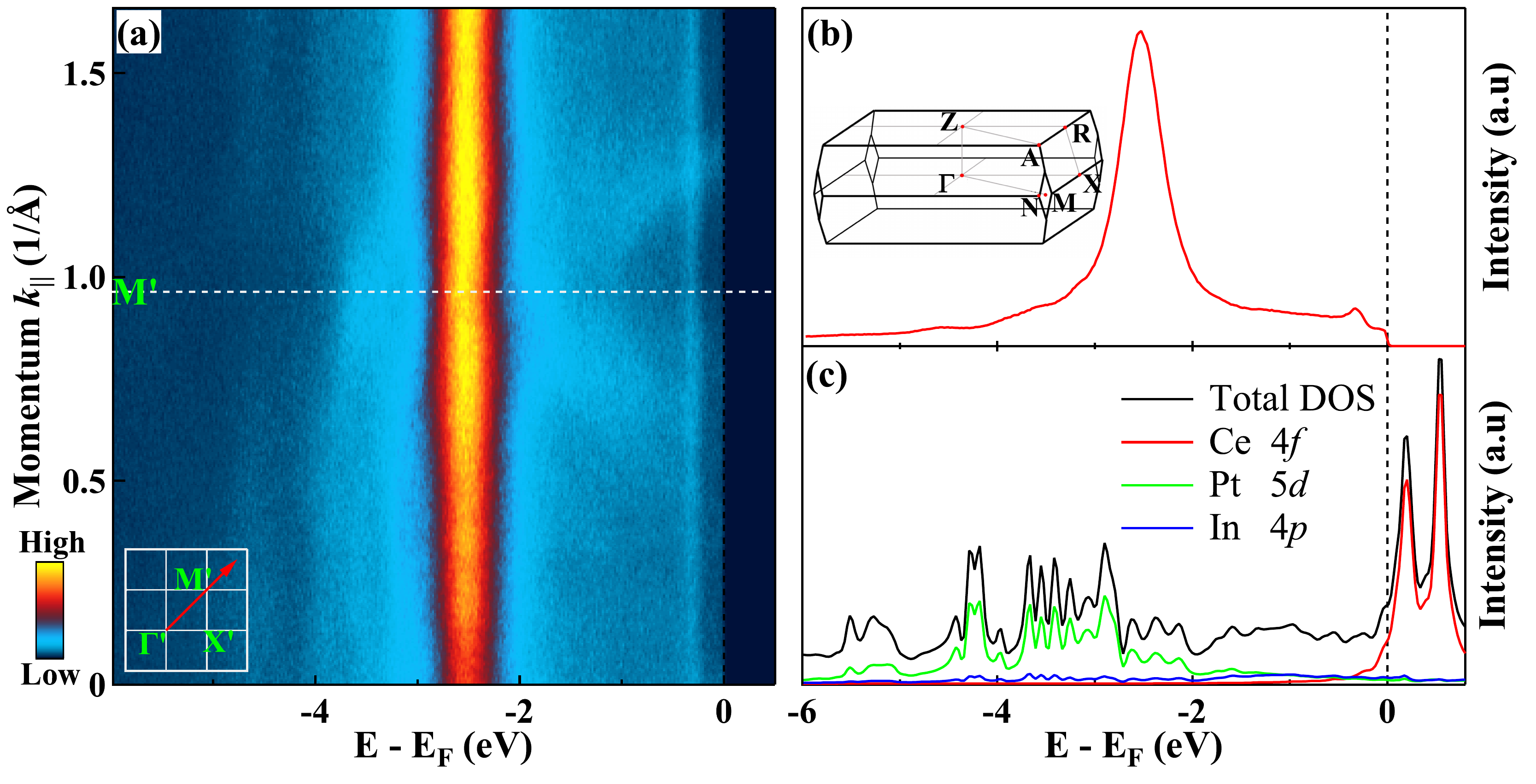}
\end{center}
\vspace*{-0.7cm}
\caption{(color online) ({\bf {a}}) {\CPI} on-resonance valence band structure at 10 K and 123 eV. ({\bf {b}}) Angle-integrated photoemission spectroscopy of the intensity plot in (a). Inset: A 3D Brillouin zone with high-symmetry momentum points (red dots) marked. The locations of $\Gamma$, $M$, $N$, and $X$ points are at the $k_z$ = 0 plane while the $Z$, $A$, and $R$ points are at $k_z$ = 2$\pi$/$c$ plane. ({\bf {c}}) Calculated Ce 4$f$, Pt 5$d$, and In 4$p$ density of state (DOS) vs.\ energy $E$.}
\end{figure}

Quantum oscillation \cite{MMAltarawneh2011} and earlier ARPES measurements at fixed low-energy ($h\nu$=21.2 eV) \cite{BShen2017} are consistent with a complex Fermi-surface (FS) topology composed of sheets with both 2D and 3D character and with relatively weak many-body renormalization effects implied from optical spectroscopy \cite{RYChen2016}. Commensurate \cite{Warren2010} or coexistence of commensurate and incommensurate \cite{HSakai2011, HSakai2012} antiferromagnetism orders were revealed by nuclear quadrupolar resonance as well as muon spin rotation/relaxation \cite{MMansson2014}. None of these earlier experiments, however, directly probed the 4$f$ electron states that are essential to the physics of this family of materials. Definitely, it is of great importance to address the nature of 4$f$ states and $f$/$d$ hybridization via on-resonance ARPES. Recent DFT/GGA calculations, which are in good agreement with ARPES measurements, suggest that the FSs near the $M$($A$) (zone corner) show a nearly 2D nature and FSs at other momenta show a strong 3D nature \cite{BShen2017}. The $k$$_z$ dispersion must be considered when analyzing and interpreting ARPES data. Here we report APRES at variable photon energies, which allows us to probe both the 4$f$ states as well as the rest of the electronic structure of {\CPI} at various $k_z$.

From these ARPES measurements on high-quality single crystals {\CPI} \cite{Tobash2012} and employing Ce 4$d$-4$f$ on-resonance spectroscopy, we have identified fine structures of spin-orbit splitting and crystal electric field (CEF) splitting of Ce 4$f$ bands. Furthermore, we have determined the $k_z$-dependent band structures at various photon energies. We find that the experimental FS topology and band structures are in good agreement with our DFT calculations.

The high-resolution ARPES experiments were performed on SIS X09LA beamline at the Swiss light Source, using a VG-SCIENTA R4000 photoelectron spectrometer. All samples were cleaved $in$ $situ$ and measured in an ultra high vacuum with a base pressure better than 4$\times$10$^{-11}$ mbar. The Ce 4$f$ state characteristics were obtained by on-resonance spectra ($h\nu$ = 123 eV) at a low temperature of 10 K. A variety of photon energies were used to investigate the 3D nature of the electron bands. An angular resolution of 0.2$^{\circ}$ was used for all measurements. High-quality single {\CPI} crystals were grown from In flux \cite{Tobash2012}.

\begin{figure}[t!]
\vspace*{-0.2cm}
\begin{center}
\includegraphics[width=0.95\columnwidth,angle=0]{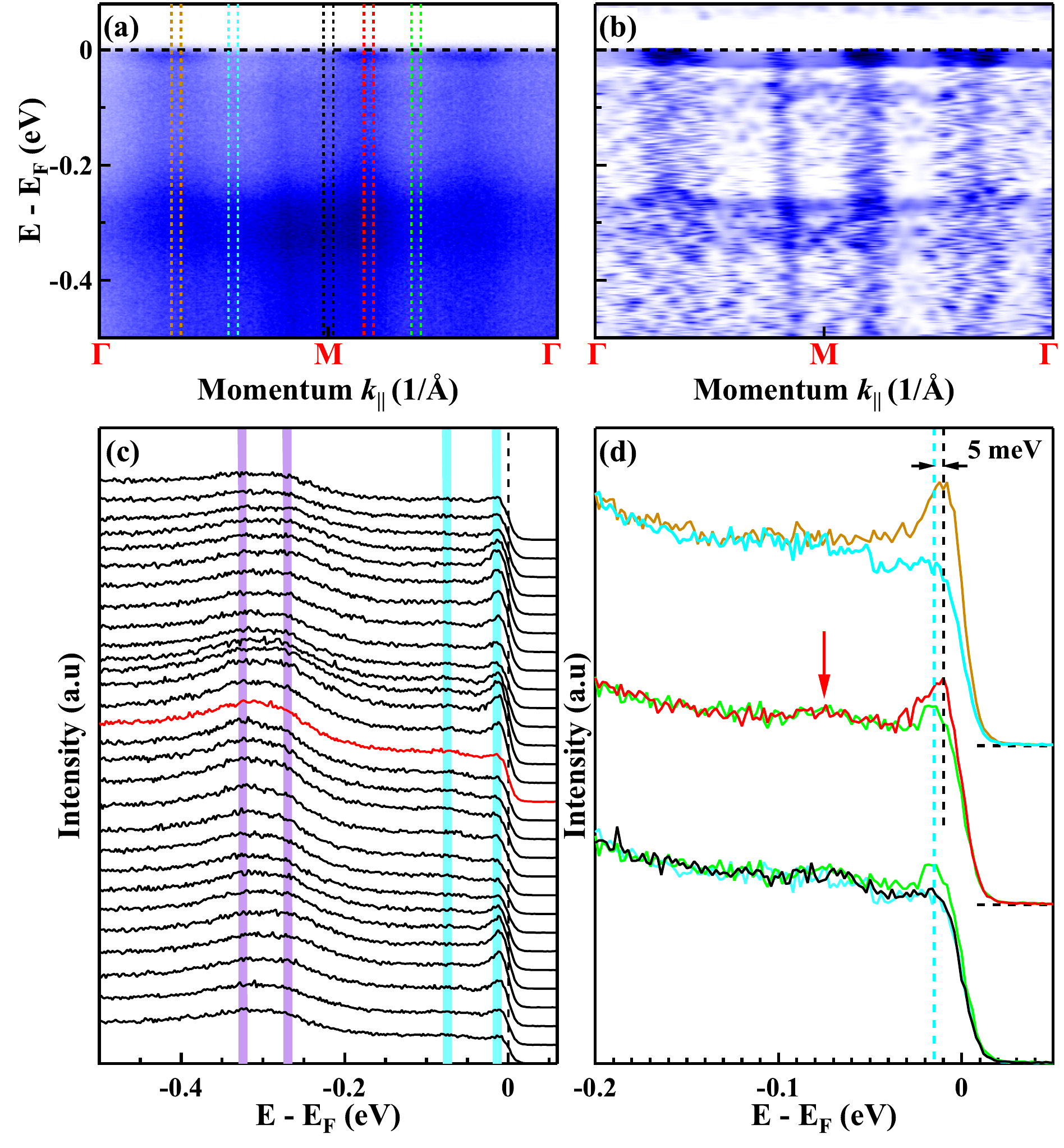}
\end{center}
\vspace*{-0.7cm}
\caption{(color online) CEF splitting of {\CPI} 4$f$ states. ({\bf {a}}) High-energy resolution of on-resonance valence band structure plot along [110] direction near Fermi energy is taken at 10 K with 123 eV. ({\bf {b}}) Measured structure reduced with the 2D second derivative to enhance the weak bands while maintaining band dispersion. ({\bf {c}}) EDCs of the spectral show in (a). The purple and aqua shadows are indicated the positions of the {$f_{5/2}^1$} and {$f_{7/2}^1$} states, respectively. The red curve represents the EDC at the M point. ({\bf {d}}) EDCs at different momentum locations in (a). The brown, cyan, black, red, and green curves are obtained by integrating the regions between two brown, cyan, black, red, and green dashed lines, as indicated in (a), respectively.
}
\end{figure}

\begin{figure*}[hbt]
\centering
\includegraphics[width=2\columnwidth,angle=0]{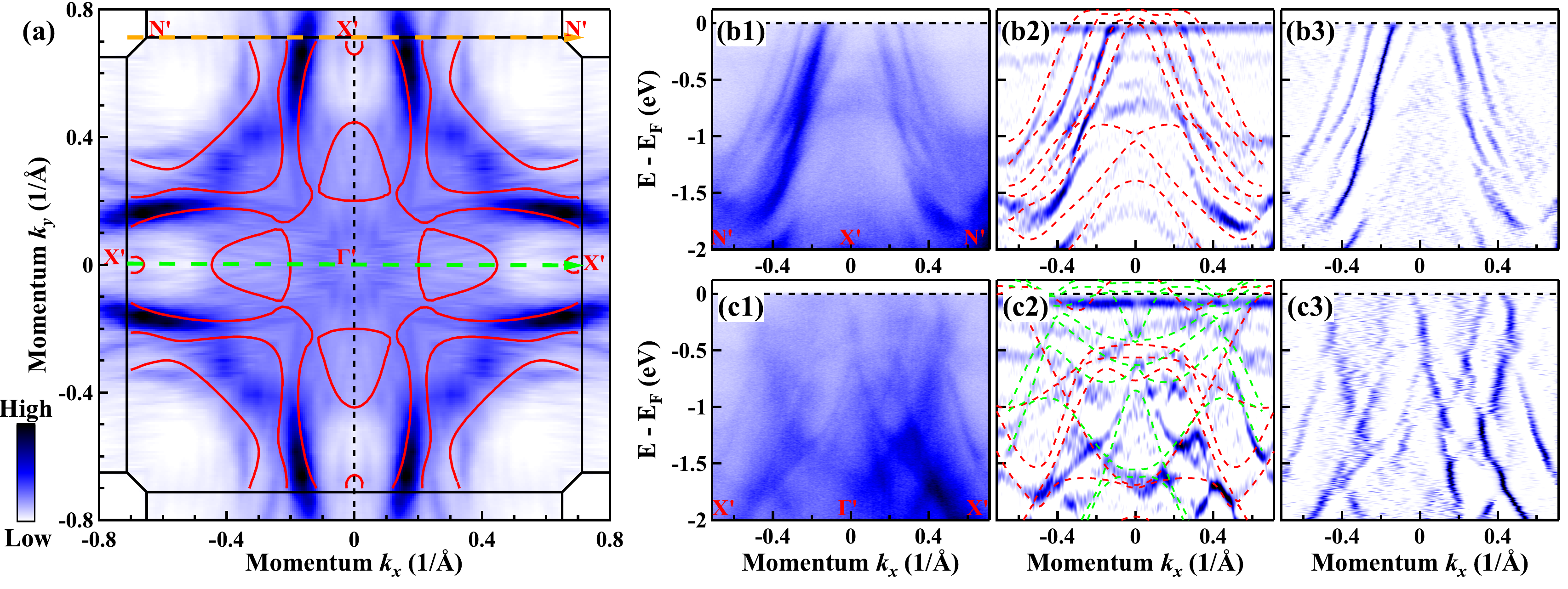}
\caption{(color online) FS and corresponding {\CPI} band structure. ({\bf {a}}) Symmetrized FS in the $k_x$-$k_y$ plane of {\CPI} measured at 10 K with 100 eV (close to $\Gamma$ point). The intensity is integrated over [-10 meV, 10 meV] energy window with respect to the Fermi level $E$$_F$. [({\bf {b1}}) - ({\bf {b3}})] Measured band structures along the  $N$$^{\prime}$--$X$$^{\prime}$--$N$$^{\prime}$ direction, indicated by the dashed yellow arrow in (a). [({\bf {c1}}) - ({\bf {c3}})] Measured band structure along  $X$$^{\prime}$-$\Gamma^{\prime}$-$X$$^{\prime}$ direction, indicated by the dashed green arrow in (a). ({\bf {b1}}) and ({\bf {c1}}) are the original data. [({\bf {b2}}) and ({\bf {c2}})] Second-derivative images from the original data with respect to energy. ({\bf {b3}}) and ({\bf {c3}}) are second-derivative images from the original data with respect to momentum. Red dashed lines in ({\bf b2}) and ({\bf c2}) represent DFT band structure. Green dashed lines represent DFT band structure along $R$-$Z$-$R$ line.}
\end{figure*}

Tuning the photon energies toward 123 eV, a strengthening of Ce 4$f$ spectral weight near the Fermi energy is expected. Figure. 1(a) shows the on-resonance ($4d\rightarrow 4f$) spectra taken from a freshly cleaved, single {\CPI} crystal as measured along the high-symmetry [110] direction at a temperature of 10 K with a total energy resolution $\sim$30 meV. The energy distribution curve (EDC) integrated over a momentum range along the measured cut appears in Fig. 1(b). A high-intensity, nondispersive structure at around $-2.5$ eV corresponds to the $f^0$ final state \cite{Fujimori2003, Fujimori2006, SPatil2016, QYChen2018A, QYChen2018B, YunZhang2018}. Weak heavy quasiparticle bands close to $E_F$ correspond to the $f^1$ final state. These low-energy peaks originate from the spin-orbit splitting of the $f^1$ final state \cite{Fujimori2003, Fujimori2006, SPatil2016, QYChen2018A, QYChen2018B, YunZhang2018, QYChen2017}. The {$f_{5/2}^1$} final state is located near $E_F$. The {$f_{7/2}^1$} final state peak is located at about $-300$ meV. The Ce 4$f$ on-resonance spectrum, even at these intensities enhanced via $4d \rightarrow 4f$ resonance, shows only limited spectral weight in the vicinity of Fermi energy.

Our DFT calculations \cite{WIEN2k,DFTdetails} suggest that most 4$f$ states are located about 0.5 eV above $E_F$, see Fig.\ 1(c). Only a very small fraction appears below $E_F$, which would explain why the intensity of $f^1$ state in {\CPI} is so low compared to the U-based heavy fermions with different $f$ occupations \cite{JDDenlinge2001,JQM_PRL2013}. The EDC shows a structure similar to the $4d \rightarrow 4f$ resonant photoemission spectra of the HF materials CeRhIn$_5$ \cite{Fujimori2003, QYChen2018A} and CeIrIn$_5$ \cite{Fujimori2003, Fujimori2006, QYChen2018B}. These spectral features observed in CeIrIn$_5$ compounds were previously understood within a single-impurity Anderson model (SIAM) framework \cite{Gunnarsson1983}, suggesting that the 4$f$ electrons are nearly localized \cite{Fujimori2003} or dominated by the localized character with a small itinerant component \cite{Fujimori2006}. While recent temperature dependence of $f$ spectral weight studies found a stronger $c-f$ electron hybridization in CeIrIn$_5$ \cite{QYChen2018B}.

According to SIAM, the $f^1$ peak intensity increases as hybridization strength increases between $f$ electrons and conduction electrons. Our ARPES measurements here reveal that the $f^1$ to $f^0$ intensity ratio is smaller for {\CPI} than for either CeIrIn$_5$ or CeRhIn$_5$ \cite{Fujimori2003}. This suggests that hybridization between $4f$ electrons and conduction electrons in {\CPI} is significantly weaker than in CeIrIn$_5$ and CeRhIn$_5$. This is consistent with prior results \cite{RYChen2016} and is in line with expectations \cite{EDBauer2010A, VASidorov2013}. Attempts to perform DFT calculations in $4f^1$ localized configuration failed to stabilize such state, providing further support to observed reduced $4f^1$ peak intensity.

We carried out higher-energy resolution ARPES measurements along the [110] direction at an overall energy resolution better than 20 meV in order to study {\CPI} 4$f$ state-related low-energy band structures. As expected, the {$f_{5/2}^1$} and {$f_{7/2}^1$} states become clearer in Fig. 2(a), and their fine structures can be found if look closely. Also, it can be found that the spectral intensities of both the {$f_{5/2}^1$} and {$f_{7/2}^1$} states show strong momentum dependence.
Figure. 2(b) shows the band structures reduced with the second derivative to enhance the weak bands while maintaining band dispersion. Four flat bands can be seen in the figure, which means that both the {$f_{5/2}^1$} and {$f_{7/2}^1$} states have undergone CEF splitting. Figure. 2(c) shows the EDCs corresponding to Fig. 2(a). The lower $f_{7/2}^1$ final state splits into two peaks. They are located at about $-325$ and $-270$ meV. The upper $f_{5/2}^1$ final-state fine structures were resolved and constitute two peaks: one at around $-75$ meV, the other at $-15$ meV. The weak, but observable, $-75$ meV feature can be seen more clearly in Fig.2(d), as indicated by a red arrow. Recently, 4$f_{7/2}^1$ and $f_{5/2}^1$ peak splitting has been detected by ARPES \cite{QYChen2018B, SPatil2016} along with inelastic neutron scattering \cite{Christianson2004, Christianson2005, Willers2010} in other Ce-based HF compounds.

In the HF system, due to the hybridization between the conduction band and the Ce 4$f$ state, it is expected that a dispersive quasiparticle band can be observed near $E_F$ at a position where the $f$ band intersects the conduction band. Figure. 2(d) display the integrated EDCs at different momentum locations in Fig. 2(a), including crossing high-intensity $f$ band regions where the conduction bands cross the Fermi level and adjacent low-intensity regions where no conduction bands cross the Fermi level. From the top and middle sets of EDCs in Fig. 2(d), we can found that the peak positions of EDCs near the $E_F$ that cross two different regions have shifted a little. The observation of weak dispersive $f$ band and locally enhanced $f$ band intensity imply the possible hybridization between $f$ band and conduction band. Hybridization occurs at where the conduction band crosses or approaches the Fermi level (brown and red lines), whereas no band hybridization is observed in other regions (cyan, green, and black lines). This indicates that dispersive and hybridized 4$f$ electron densities contribute to bonding and FS formation. This has been suggested, too, by quantum oscillation measurements \cite{MMAltarawneh2011}. The hybridization between $f_{5/2}^1$ electron and conduction bands also has been reported in other Ce-based heavy fermions \cite{Fujimori2006, QYChen2018A, QYChen2018B, YunZhang2018, QYChen2017, Fujimori2016}. As shown in Fig. 2(d), the energy dispersion of the hybridized band is about 5 meV, which is much smaller than that of other HF compounds. The energy dispersion of the hybridized quasiparticle bands is more than 10 meV for CeIrIn$_5$ \cite{Fujimori2006} and CeCoIn$_5$ \cite{QYChen2017}. According to the periodic Anderson model, the stronger the hybridization between the $f$ electrons and the conduction electrons, the larger the energy dispersion of the hybridized quasiparticle band. Again, this means the hybridization between $f$ bands and conduction bands in {\CPI} is weak compared to CeIrIn$_5$ and CeCoIn$_5$. Comparison of theoretical band structures and Fermi surface contours for {\CPI} and LaPt$_2$In$_7$ also support this point (see Fig. S1 of the supplemental material \cite{SuppMaterial} for more details).

\begin{figure*}[hbt]
\centering
\includegraphics[width=2\columnwidth,angle=0]{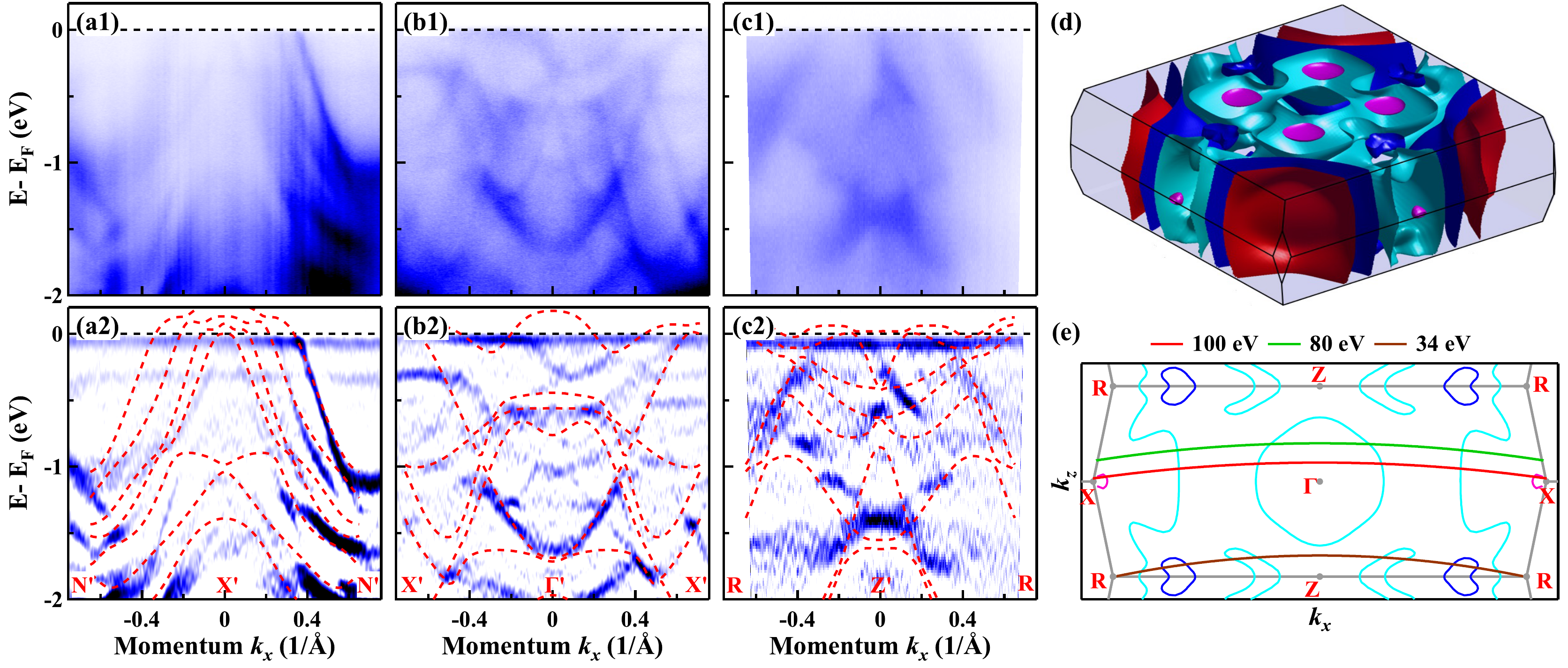}
\caption{(color online) Band structure of {\CPI} at 10 K. [({\bf {a1}}) and ({\bf {a2}})] Original photoemission image and corresponding EDC second derivative image measured along the $N$$^{\prime}$-$X$$^{\prime}$-$N$$^{\prime}$ direction with 80-eV photon energy. [({\bf {b1}}) and ({\bf {c1}})] Photonemission data taken along $X$$^{\prime}$--$\Gamma$$^{\prime}$--$X$$^{\prime}$ direction with 80- and 34-eV photon energies, respectively. [({\bf {b2}}) and ({\bf {c2}})] Corresponding EDC second derivative images of ({\bf {b1}}) and ({\bf {c1}}), respectively. Red dashed lines in ({\bf a2}), ({\bf b2}), and ({\bf c2}) represent DFT band structure. ({\bf {d}}) Calculated FS of {\CPI}. The 3D FS is presented in the body-centered tetragonal Brillouin zone, in which one unit cell contains one Ce atom. ({\bf {e}}) DFT-calculated Fermi surface contours in the $\Gamma$$XRZ$ plane ($k_x$-$k_z$). The colors used for different Fermi surfaces are consistent with ({\bf {d}}). The red, green, and brown arcs indicate the final-state arcs for 100-,80- and 34-eV photon energies, respectively.}
\end{figure*}

Figure. 3(a) shows high-resolution FS mapping of {\CPI} measured at 10 K using a constant photon energy of 100 eV. This corresponds to a cut close to the $\Gamma$ point ($k_z$$\sim$ 0.2 $\times$ 2$\pi$/$c$), estimated based on an inner potential of 11 eV \cite{SuppMaterial}. FS contours are drawn with false colors. To compare to DFT calculations, the calculated FS contours were overlaid on top of the experimental FS. Here, the FS contour at $k_z$ = 0 is selected to capture features around $X$ point, because the end of the final-state arc of 100 eV has already touched the $X$ point. Multiple FS sheets were revealed in the momentum space covered. Three electron pockets can be found around the zone corners, and a small hole pocket can be found located around the $X$ point, agrees well with the band structure predictions. Figure. 3(b1) and 3(c1) show the detailed band structures along the two high-symmetry directions as indicated by dashed yellow and dashed green arrows in Fig. 3(a).

The EDC bands [Figs.\ 3(b2) and 3(c2)] and momentum distribution curve (MDC) bands [Figs.\ 3(b3) and 3(c3)] were extracted by taking the second derivatives of the original ARPES data [shown in Figs.\ 3(b1) and 3(c1)] with respect to energy and momentum, respectively. Possible band dispersions become thereby easier to distinguish. Our calculated band structures (red-dashed lines) were overlaid onto the EDC band [Figs.\ 3(b2) and 3(c2)]. It can be found that the measured and computed band structure are in good agreement, especially along the $N$$^{\prime}$--$X$$^{\prime}$--$N$$^{\prime}$ direction. Along the $N$$^{\prime}$--$X$$^{\prime}$--$N$$^{\prime}$ direction, where the FSs are nearly 2D [Fig. 4(d)] (see Fig. S2 of the supplemental material for more details \cite{SuppMaterial}), experimental data and calculated results are highly consistent [Fig.\ 3(b2)]. While along the $X$$^{\prime}$--$\Gamma$$^{\prime}$--$X$$^{\prime}$  direction, where the FSs show strong 3D nature [Figs. 4(d)], the experimental results are more complex than theoretical calculations (red dashed lines) [Figs.\ 3(c2)]. It has been pointed out that there is a very significant $k_z$ broadening in Ce-based HF materials \cite{QYChen2018A}. As we know, the $k_z$ broadening has a great influence on the 3D FS, but little influence on the 2D FS. Dispersions along $R$$^{\prime}$--$Z$$^{\prime}$--$R$$^{\prime}$ direction (green dashed lines, $k_z$= 2$\pi$/$c$) were also overlaid on top of Fig. 3(c2). It can be found that some of the other dispersions are consistent with the dispersion from $R$--$Z$--$R$. All these results strongly suggest that the $k_z$ broadening should carefully be taken into account when discussing the origin of photon energy-dependent spectral features.

Figure. 4 shows ARPES spectra measured at 10 K along the N$^{\prime}$--X$^{\prime}$--N$^{\prime}$ and $X$$^{\prime}$--$\Gamma$$^{\prime}$--$X$$^{\prime}$ directions. The energy resolutions were set at 22 and 15 meV for 80- and 34-eV photon energies, respectively. Figure. 4(a1) and 4(a2) show original and EDC band structure taken along the $N$$^{\prime}$--$X$$^{\prime}$--$N$$^{\prime}$ direction with 80-eV photon energy, respectively. Obviously, at the zone boundary, although the measured $k_z$ positions are different, the band structures for 80 and 100 eV [Fig. 3(b1)] are very similar, four bands crossing or approaching the Fermi level. Figure. 4(b1) and 4(c1) show original band structure measurements taken along $X$$^{\prime}$--$\Gamma$$^{\prime}$--$X$$^{\prime}$ direction with 80- and 34-eV photon energies, respectively. Figure. 4(b2) and 4(c2) are EDC bands from the original data. The band structures show significant photon-energy $h\nu$ dependence, which indicates a very strong $k_z$ dispersion. Thus, our ARPES measurements are consistent with a strong 3D FS topology, in good agreement with the DFT calculation [Fig. 4(d)]. And in the supplemental material \cite{SuppMaterial}, we display the detailed calculated FS contours of {\CPI} at different $k_z$s. Similarly to earlier results from another research group \cite{BShen2017}, our experimental and theoretical results show that the FSs near the $M$($A$) (zone corner) show a good 2D nature and FSs at other momenta show a significant 3D nature.

In conclusion, {\CPI} FS topology and low-energy band structures were investigated at 10 K using high-resolution ARPES. We directly observed heavy 4$f$-derived quasiparticle bands via on-resonance ARPES. Our study show that the hybridization between 4$f$ electrons and conduction bands are weak, and the hybridized $4f$ electrons contribute to bonding and FS formation. The CEF splitting of $4f_{5/2}^1$ and $4f_{7/2}^1$ states was directly observed via on-resonant measurements. We also confirm that the FS has a strong 3D topology, well supported by DFT calculations. These findings provide key insight into understanding the electronic structure of an unconventional superconducting Ce-based HF.

We acknowledge helpful discussions with Professor Jian-Xin Zhu. This work was supported by the National Natural Science Foundation of China (Grant No.\ 11574402),ZDXKFZ (Grant No.\ XKFZ201703), and the Innovation-driven Plan in Central South University (Grant No.\ 2016CXS032). J.R.\ and P.M.O.\ acknowledge support through the Swedish Research Council (VR) and the Swedish National Infrastructure for Computing (SNIC), for computing time on computer cluster Triolith at the NSC center Link{\"o}ping. Y.S. acknowledges the support from the Swedish Research Council (VR) through a Starting Grant (Dnr. 2017-05078). O.T. acknowledges support from the Swedish Research Council (VR) and the Knut and Alice Wallenberg foundation. M.M. is partly supported by a Marie Sklodowska-Curie Action, International Career Grant through the European Commission and Swedish Research Council (VR), Grant No. INCA-2014-6426, as well as a VR neutron project grant (BIFROST, Dnr. 2016-06955). Further support was also granted by the Carl Tryggers Foundation for Scientific Research (Grant No.\ CTS-16:324 and No.\ CTS-17:325). Work at Los Alamos was performed under the auspices of the U.S. Department of Energy, Office of Basic Energy Sciences, Division of Materials Sciences and
Engineering.

{}
\end{document}